\documentclass{ar-1col}

\jname{Xxxx. Xxx. Xxx. Xxx.}
\jvol{AA}
\jyear{YYYY}
\doi{10.1146/((please add article doi))}

\usepackage{wrapfig}
\usepackage{amsbsy}
\usepackage{enumitem}
\usepackage{amssymb}
\usepackage{natbib}       
\usepackage{graphicx}

\include{psfig}

\newcommand{\Ivec}{\mbb{I}}

\newcommand{\nvec}{\mbb{n}}

\newcommand{\beqar}{\begin{eqnarray}}
\newcommand{\eeqar}{\end{eqnarray}}

\newcommand{\eq}{\equiv}

\newcommand{\nl}{\newline}
\newcommand{\bc}{\begin{center}}
\newcommand{\ec}{\end{center}}
\newcommand{\bslide}{\begin{slide}}
\newcommand{\eslide}{\end{slide}}

\newcommand{\be}{\begin{equation}}
\newcommand{\ee}{\end{equation}}
\newcommand{\bd}{\begin{displaymath}}
\newcommand{\ed}{\end{displaymath}}
\newcommand{\bea}{\begin{eqnarray}}
\newcommand{\eea}{\end{eqnarray}}
\newcommand{\bda}{\begin{eqnarray*}}
\newcommand{\eda}{\end{eqnarray*}}
\newcommand{\ba}{\begin{array}}
\newcommand{\ea}{\end{array}}
\newcommand{\bit}{\begin{itemize}}
\newcommand{\eit}{\end{itemize}}
\newcommand{\ben}{\begin{enumerate}}
\newcommand{\een}{\end{enumerate}}
\newcommand{\bt}{\begin{tabular}}
\newcommand{\et}{\end{tabular}}
\newcommand{\beab}{\begin{abstract}}
\newcommand{\eab}{\end{abstract}}

\newcommand{\bm}{\boldmath}

\newcommand{\pa}{\partial}

\newcommand{\pqdr}[1]{\left[#1\right]}
\newcommand{\pgrf}[1]{\left\{#1\right\}}

\newcommand{\mean}[1]{\left\langle #1 \right\rangle}
\newcommand{\del}{\delta}

\newcommand{\Om}{\Omega}

\newcommand{\kaj}{\kappa_j}
\newcommand{\kak}{\kappa_k}
\newcommand{\kal}{\kappa_l}
\newcommand{\ka}{\kappa}
\newcommand{\Ka}{\mbox{\boldmath$\kappa$}}
\newcommand{\Kap}{\Ka'}

\newcommand{\sumkp}{\sum_{\Kap}}

\newcommand{\sumband}{\sum_{\ka\le|\Ka| < \ka+1}}

\newcommand{\ukkp}{\hat{u}_k(\Ka')}
\newcommand{\ulkkp}{\hat{u}_l(\Ka-\Ka')}

\newcommand{\huj}{\hat{u}_j}

\newcommand{\hujs}{\huj^\ast}

\newcommand{\hujsk}{\hujs(\Ka)}

\newcommand{\mbb}[1]{\mbox{\boldmath$#1$}}

\newcommand{\uvec}{\mbb{u}}

\newcommand{\Uvec}{\mbb{U}}
\newcommand{\Vvec}{\mbb{V}}

\newcommand{\vorts}{\mbb{\Om}_s}

\newcommand{\xvec}{\mbb{x}}

\newcommand{\fvec}{\mbb{f}}

\newcommand{\Pjk}{P_{jk}}

\newcommand{\Pjktensor}{\del_{jk}-\frac{\kaj\kak}{\ka^2}}

\newcommand{\Tk}{T(\ka)}

\newcommand{\hfj}{\hat{f}_j}

\newcommand{\Real}[1]{\Re\pgrf{#1}}
\newcommand{\Imag}[1]{\Im\pgrf{#1}}

\newcommand{\noi}{\noindent}
\newcommand{\bdm}{\begin{displaymath}}
\newcommand{\edm}{\end{displaymath}}

\newcommand{\bS}{$ \textbf{S}\, $}   
\newcommand{\bSinv}{$ \textbf{S}^{-1}\, $}
\newcommand{\BS}{ \textbf{S}}
\newcommand{\BEE}{ \textbf{E}}
\newcommand{\BI}{ \textbf{I}}
\newcommand{\BOM}{\mbb{\Omega}}

\definecolor{g-blue}{rgb}{0.83,0.95,1}
\definecolor{g-yellow}{rgb}{1,1,0.7}
\definecolor{g-green}{rgb}{0.9,1,0.9}
\definecolor{green}{rgb}{0,0.6,0}
\definecolor{cyan}{rgb}{0,0.7,0.7}
\definecolor{black}{rgb}{0,0,0}
\definecolor{grey}{rgb}{0.4 ,0.4 ,0.4 }

\def\black#1{\textcolor{black}{#1}}

\def\red#1{\textcolor{red}{#1}}

\usepackage{calrsfs}
\newcommand{\BC}[1]{\bm{\mathcal{#1}}}
\def\timeofday{\count50 = \number\time
\count51 = \number\time \count53 = \number\time \divide\count50 by
60 \multiply\count50 by 60 \divide\count51 by 60 \advance\count53
by -\count50 \ifnum\count53>9 \number\count51 :\number\count53
\else \number\count51 :0\number\count53 \fi}
\def\be{\begin{equation}}\def\ee{\end{equation}}
\def\bea{\begin{eqnarray}}\def\eea{\end{eqnarray}}
\def\bse{\begin{subequations}}\def\ese{\end{subequations}}
\newcommand{\BE}[1]{\begin{equation}\label{#1}}
\newcommand{\BEA}[1]{\begin{eqnarray}\label{#1}}
\newcommand{\BSE}[1]{\begin{subequations}\label{#1}}

\usepackage{hyperref}
\hypersetup{
     pdfborder = {0 0 0},
  colorlinks   = true, 
  urlcolor     = blue, 
  linkcolor    = blue, 
  citecolor    = red   
}
\begin{document}

\title{ DNS of Turbulent Flows Laden with Droplets or Bubbles}
\author{S. Elghobashi
\affil{Mechanical and Aerospace Engineering Department,}
\affil{University of California, Irvine, California 92689}}
%
\setcounter{tocdepth}{3}
\setcounter{secnumdepth}{3}
\begin{abstract}
\red{\large  \bf \today  - \timeofday}\\
\black{This review focuses on Direct numerical simulations (DNS) of turbulent flows laden with droplets or bubbles. DNS of these flows are more challenging than those of flows laden with solid particles due to the surface deformation in the former. The classification of the discussed numerical methods is based on whether the initial diameter of the bubble/droplet is smaller or larger than the Kolmogorov length scale, and whether the instantaneous surface deformation is fully resolved or obtained via a phenomenological model.  Numerical methods that account for the breakup of single droplet/bubble as well as multiple droplet/bubble in canonical turbulent flows are also discussed.
}
\end{abstract}

\maketitle

\newpage
\section{INTRODUCTION}
\label{sec:intro}
Turbulent flows laden with liquid droplets or gas/vapor bubbles (also known as turbulent dispersed multiphase flows)   are ubiquitous in nature and engineering applications.
In nature, examples include rain, waterfall mists,   air bubbles in the upper ocean, and vapor bubbles in geysers.  Engineering applications include liquid fuel sprays in all types of combustion engines, paint sprays, spray drying in the pharmaceutical industry as well as   food processing, and water vapor bubbles in nuclear reactor cooling systems or those created by cavitation  in the wakes of ship propellers, just to list a few.
\par
Direct numerical simulations (DNS)  of turbulent flows laden with droplets or bubbles are more challenging than DNS of solid particle-laden turbulent flows  due to 
the shape deformation of the dispersed phase in the former.
Accurate prediction of the deformation of the interface between the dispersed and continuous phases
 requires proper accounting of the effects of surface tension and the different viscosities and densities of the two phases in the governing equations of motion. \black{A dimensionless measure of the ability of the carrier fluid motion to deform the immersed droplets or bubbles is the Weber number, $We$, which is the ratio of the inertial forces to surface tension forces. Qualitatively,  large $We$ values enhance the deformability whereas lower values reduce it.
 Another dimensionless measure is Ohnesorge number, $Oh$, which is the ratio of viscous to surface tension effects, and is related to $We$ and Reynolds number of the bubble/droplet according to
$Oh= \sqrt{We}/Re$.  For a fixed $Re$ the effects of changing  $Oh$ are similar to those of changing $We$.  }
\par
The size of  the droplets or bubbles  in the reviewed studies
ranges from smaller than to larger than the Kolmogorov length scale.
 In all the DNS studies reviewed here the governing conservation equations of the interacting  fluid phases are solved on a fixed (Cartesian) grid. In other words, methods that use {\em interface-fitting adaptive grids} are not considered in this review \black{ since these  methods considered only non-turbulent flows.}
 \par \black{Since current supercomputers allow DNS to resolve only the turbulent fluid motion of length scales equal or greater than the Kolmogorov length scale,
 then it is not possible at present to fully  resolve the motion of  dispersed deformable sub-Kolmogorov scale droplets or bubbles.
 In order to overcome this difficulty, phenomenological models are used to compute the deformation of the dispersed phase as will be discussed in
  sections \ref{sec:deformicrob} and \ref{sec:drop<eta}.  }
  \par  \black{For deformable bubbles or droplets whose size is larger than the Kolmogorov length scale,
  the resolved shape and motion of the interface between the two phases are computed via one of the following {\em three} approaches:}
 \begin{enumerate}[label=\Alph*.]
\item	{\em Tracking points: }
\nl Here the interface is marked by points that are advected by the flow as in the front tracking method of \cite{tryggv1992} and \cite{tryggv2001}.
\item	{\em  Tracking scalar functions:}
\nl Each of the following four numerical methods has its tracking  function.
\begin{enumerate}[label=\arabic*.]
\item Volume of fluid (VOF), where the function is the volume fraction of the local phase on either side of the interface (\cite{scardovelli1999}).
\item  Level Set (LS), where the function is the signed distance function representing the shortest distance from the interface (\cite{sussman1994, osher2001}) or its hyperbolic-tangent version (\cite{desj2008}).
\item Lattice Boltzmann (LB), where the function  $f^n_i(\xvec,t)$ is the probability density function of finding a fluid particle of each fluid phase $n$  at position $ \xvec$,  time $t$,  and moving in the direction $i$ of one of the discretized lattice velocity directions.
    The physical properties of the fluid such as the density or momentum are defined as moments of  $f^n_i(\xvec,t)$. In the discretized LB method (LBM) the moments are evaluated by quadrature summation over all $i$.
The interface between two phases  is modeled by adding an extra force to the LB equilibrium velocity to represent the microscopic interaction
 between the two phases (e.g. surface tension or diffusivity) (\cite{shan1993}).
\item
 Phase Field model (PFM), where the function is the {\em scalar phase field}, $\phi(\xvec,t)$, also known as {\em order parameter},
 which represents one of the physical properties (e.g.   molar concentration) of a binary fluid mixture.
 $\phi(\xvec,t)$ is mostly uniform in the bulk phases and varies smoothly over a  {\em diffuse} finite-thickness interfacial layer.
 The transport of $\phi(\xvec,t)$ is governed by the Cahn-Hilliard equation (\cite{cahn1959}) which accounts for the advection of $\phi(\xvec,t)$   by the fluid velocity and the diffusion which equals
   $\nabla \cdot [M(\phi) \nabla \mu_\phi]$,
  where  $M(\phi)$ is the fluid mobility, and  $\mu(\phi)$ is the chemical potential
  which is defined in terms of the free energy $f(\phi)$ of the fluid.
 In contrast to the above three  methods, here the surface tension forces are replaced by  a continuum model of
 $f(\phi)$ (\cite{jacqmin1999}).
Accordingly, the Navier-Stokes equations are modified by adding  the forcing function ($ \mu \nabla \phi$) to represent the surface tension forces
(\cite{gurtin1996}).
\nl
Recently, a hybrid LBM-PFM was used to simulate  the dispersion of liquid droplets in isotropic turbulence  (\cite{komrakova2015}) as will be discussed in section \ref{sec:LBM-PFM}.
%
%
\end{enumerate}
\item  \black{{\em Immersed boundary method (IBM) with interaction potential model (IPM):}
This hybrid  IBM-IPM is a recently developed approach (\cite{spandan2017})
 that couples the immersed boundary method (IBM) with a phenomenological {\em interaction potential model} (IPM) to simulate deformable droplets or bubbles in a turbulent  flow. The dynamics of the interface deformation is modeled via a 3D spring network distributed over the surface of the immersed droplet (\cite{tullio2016}). The IPM  is based on the principle of minimum potential energy where the total potential energy depends on the extent of deformation of the spring network. Modeling the spring network requires computing ad-hoc elastic constants which is done through a reverse-engineered approach.
The IBM enforces the boundary conditions at the interface (e.g. the no-slip).  A moving-least-squares (MLS) approximation (\cite{vanella2009}) is used to reconstruct the solution in the vicinity of the immersed surface and to convert the Lagrangian forcing back to the Eulerian grid.  MLS ensures constructing uniform Lagrangian grid elements on the immersed surface  as it deforms.
\nl \cite{spandan2018} used IBM-IPM to study the  deformation of bubbles dispersed in a turbulent Taylor-Couette flow and the effect of their deformation on drag reduction.
This study will be discussed in section \ref{sec:bub-TC}.
}
\end{enumerate}
\par \cite{anderson1998}  provides a historical review of the early studies by Poisson, Maxwell, Gibbs, Rayleigh and van der Waals on modeling the interface between two immiscible fluids.
\par  When necessary the numerical methods will be briefly discussed.
 \black{However, the focus here is not on the fine details of the different numerical algorithms but rather on the contributions of the different methods to advancing our understanding of the physics of the interactions between turbulence and droplets or bubbles. }
This review intends to complement the  recent reviews of turbulent  dispersed multiphase flows by
 \cite{bala-eaton2010}, \black{ bubble-laden turbulent flows by} \cite {tryggvason2013},  and simulation methods of particulate flows by \cite{maxey2017}.
\par
 The article proceeds as follows.
 Section 2 discusses DNS of bubble-laden turbulent flows,
 Section 3  discusses DNS of droplet-laden turbulent flows, and  Section 4 provides concluding remarks.
\vspace{-3mm}
\subsection{Glossary}
\black{The following is a list of the acronyms used in this article:
\nl EL: Eulerian Lagrangian
\nl FTM: Front Tracking Method
\nl IBM: Immersed Boundary Method
\nl IPM: Interaction Potential Model
\nl NS: Navier-Stokes
\nl PFM: Phase Field Model
\nl TF: Two-Fluid
\nl TKE: Turbulence Kinetic Energy
\nl VOF: Volume Of Fluid}
\vspace{-3mm}
\section{DNS OF BUBBLE-LADEN TURBULENT FLOWS}
The word `bubbles' in this article refers to either {\em gas bubbles or vapor bubbles}
  since  the mechanical aspects of their motion in liquid are the same
   except for the stronger effects of added mass in the latter ({\cite{prosper2017}).
\par
   The length scale $d$  will be used to denote the maximum size of the bubble. Thus, for a {\em spherical bubble}
   the length $d$  equals the diameter. For an {\em ellipsoidal bubble}, $d$ equals the length of the
   major axis. In the following sections, the discussion will consider bubbles whose size $d$ is smaller than the Kolmogorov length scale, $\eta$, as well as bubbles with $d> \eta$.
Figure \ref{fig:diag} shows a list of the authors whose papers are reviewed in this article.
\subsection{Bubbles of size smaller than the Kolmogorov length scale,  $d < \eta$}
Bubbles with $d < \eta$ are generally referred to as microbubbles (\cite{madavan84} and \cite{oleg-se-98}).
 DNS of microbubble-laden turbulence can be performed using the two-fluid (TF) approach or the Eulerian-Lagrangian (EL) approach. The EL approach is based on the point-particle assumption (\cite{se-prosper2009}, \cite{bala-eaton2010}).
\subsubsection{Nondeformable spherical bubbles with $d< \eta$}
\begin{figure}
\vspace{-5mm}
{
\includegraphics[width=5.in]{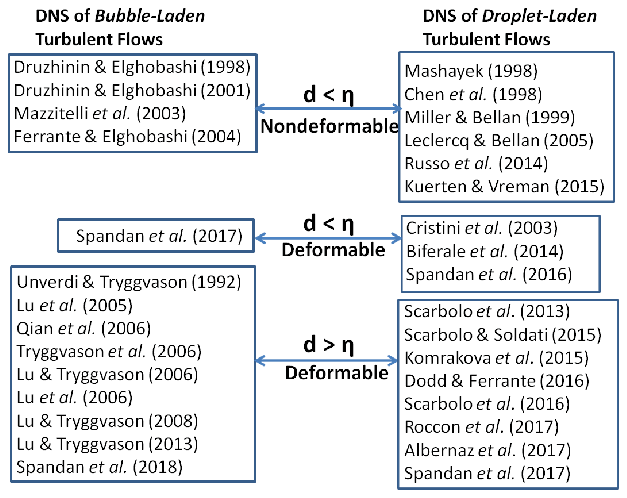}
}
\caption{Reviewed DNS studies of bubble-laden and droplet-laden turbulent flows}
\label{fig:diag}
\end{figure}
The equation of motion of  a  nondeformable  spherical microbubble in a turbulent flow has been derived
by \cite{oleg-se-98}, following the approach of \cite{max83}, for the conditions of $d < \eta$,  neglecting the Basset and lift forces, and assuming
 $\rho_b << \rho_f$,  where   $\rho_b$ and $\rho_f$ are the densities of the bubble gas and surrounding fluid (liquid) respectively, as:
\be
\label{eqbm1}
\frac{d V_i}{dt}= 3 \frac{D U_i}{Dt} + \frac{1}{\tau_b} \
(U_{i,\xvec_b}  - \ V_i + W \delta_{iz}) \ ,
\ee
where the bubble response time $\tau_b$
and terminal velocity $W$ are defined as
\be
\label{taub}
\tau_b = \frac{d^2}{36 \nu} \ ; \ \ \ W = 2 \tau_b \ \rm{g} \ .
\ee
In the above equations, $V_i$ and $U_i$   are respectively the instantaneous components of the bubble velocity
and the  carrier fluid  velocity  in the three coordinate directions. $\xvec_b$ is the position  of the bubble center,
 $\frac{D }{Dt}$ is the Lagrangian derivative, $\rm{g}$ is the gravitational acceleration, and $\nu$ is the kinematic viscosity of the carrier fluid.
 Eq.(\ref{eqbm1}) is valid for the condition  $d< \eta$  which simultaneously {\em necessitates} that $\tau_b < \frac{\tau_K}{36} $ according to Eq.(\ref{taub}), where $\tau_K$
 is the Kolmogorov time scale.  This $\tau_b$ condition  for the bubble does not have a corresponding restriction
 for the solid particle  when using the equation of \cite{max83} where the particle response time, $\tau_p$, can be larger than $\tau_K$.
 \nl \cite{oleg-se-98} derived the TF equations for a bubble-laden turbulent
 flow by spatially averaging the governing equations of the carrier fluid and the bubble phase over a scale of the order of $\eta$, which is much larger than the bubble diameter. They used the TF equations to predict a bubble-laden Taylor-Green vortex flow and decaying isotropic turbulence with two-way coupling.  The same authors used the TF equations in  DNS to study a 3D bubble-laden  spatially developing mixing layer with two-way coupling \cite{oleg-se-jfm}.
 \par
 \cite{ferrante-jfm2007} performed DNS using the TF equations to study the  effects of microbubbles on the vorticity
dynamics in a Taylor-Green vortex flow. The
results show that bubbles with a volume fraction $\sim 10^{-2}$ enhance the decay rate
of the vorticity at the center of the vortex. This is due to bubble clustering in the vortex core which creates a positive
velocity divergence. The vorticity transport equation shows that this positive $\nabla \cdot \Uvec$ enhances the local vorticity decay.
\par
 \cite{ferrante-jfm04} performed DNS of a microbubble-laden spatially developing  turbulent  boundary layer over a flat plate at $Re_{\theta}=1430$ to study drag reduction.
They included the bubble lift force $\pqdr{(\Uvec-\Vvec) \mbb{\times} \vorts}_i$ according to \cite{auton87} and \cite{ah} in Eq.(\ref{eqbm1}), where $\vorts$ is the local vorticity vector. The lift force  was needed for the bubble trajectories as they pass through the viscous  sublayer where  the fluid mean velocity gradient is maximum.
The bubble diameter was $d_b=62\,\mu\rm{m}$; $d^+_b=2.4$ in wall units, and volume fraction $\Phi_v$ varied from $0.001\,\rm{to} \  0.02$.
\nl
The authors concluded that the presence of bubbles in the  boundary layer results in a {\em local}
  positive divergence  of the fluid
 velocity, {\small $\nabla \cdot \Uvec > 0$}, creating a positive
 mean
 velocity normal to (and away from) the wall
 which, in turn, reduces the mean streamwise velocity and
 displaces the quasi-streamwise longitudinal vortical
 structures away from the wall.  This displacement has two main
 effects:  {\bf 1.}
 it increases the spanwise gaps between the wall streaks associated with
  the sweep events and reduces the streamwise velocity in these
  streaks,
thus reducing the skin friction by up to $20.2\%$ for
$\Phi_v=0.02$, and
  {\bf 2.}  it moves the location of peak Reynolds stress production
  away from the wall to a zone
  of a smaller transverse gradient of the mean streamwise velocity
   (i.e. smaller mean shear), thus reducing
  the production rate of turbulence kinetic energy and enstrophy."
\par
  The above described  drag reduction mechanism   applies  for {\em nondeformable microbubbles with $d<\eta$}.
We will discuss later in section \ref{sec:deformicrob} that drag reduction can be also be realized by {\em deformable microbubbles} as well as
 {\em deformable large bubbles with $d>\eta$} in section \ref{sec:deformb}.
\par \black{ \cite{Lohse2003} performed DNS of  microbubble-laden isotropic turbulence using the point particle approach to study the two-way coupling effects, especially that of the lift force discussed above.
They used $144000$ bubbles with $d\sim 120-250\, \mu m$  and volumetric fraction $\Phi_v =0.016$.
However, they applied artificial forcing to the turbulence kinetic energy (TKE) spectrum, $E(k,t)$, at small wavenumbers to create stationary turbulence at a fixed $Re_{\lambda}=62$. This forcing `camouflages' the true two-way coupling effects of the bubbles on the flow  and thus {\em no correct} conclusion can be made about  these effects.
The {\em camouflage} is explained by the  spectral transport equation of  $E(k,t)$:
\be
\frac{d E(k,t)}{dt}=   T(k,t)-{\varepsilon(k,t)}+ {\Psi_b(k,t)} + F(k,t) \ ,
\label{eq:spectr}
\ee
where the terms on the RHS are respectively the transfer rate of TKE  at wavenumber $k$, the dissipation rate, the bubbles two-way coupling rate, and the artificial forcing rate. The instantaneous two-way coupling and transfer rates (after omitting the $t$)  are:
\be
   {\Psi_b(k)} = - \sumband \Real{\mean{\hujs(\Ka)\hfj(\Ka)}},
\label{eq:psik}
\ee
\be
    \Tk= \sumband \kal \Pjk(\Ka)
    \Imag{\sumkp \mean{\ukkp \ulkkp \hujsk}},
\label{eq:tk}
\ee
where $\Real{}$ and $\Imag{}$ denote the real
and imaginary parts, and
\be
    \Pjk(\Ka)\eq\Pjktensor
\ee
is the projection tensor, and $\del_{jk}$ is Kronecker delta.
  $\hfj(\Ka)$ is the Fourier coefficient of the force $f_j$  imparted by the bubbles on the surrounding fluid,
and $^*$  denotes the complex conjugate.
 The RHS of both  (\ref{eq:psik})  and (\ref{eq:tk}) create nonlinear triadic interactions involving {\em all} wavenumbers of  $E(k,t)$  including the small wavenumbers where $F(k,t)$ is applied  (\cite{ferrante03}).
 These triadic interactions are responsible for {\em signaling} the effects of the perturbations created by
 the microbubbles at high wavenumbers to the large scales (small wavenumbers) (\cite{se93}).
Therefore, artificially increasing $E(k,t)$  at small wavenumbers  by $ F(k,t)$  {\em opposes} the two-way coupling effects of the microbubbles. Furthermore, by definition,  $ \int_{k_{min}}^{k_{max}}\frac{d E(k, t) dk}{dt} = \frac{d E}{dt}=0$ for a forced stationary turbulence and thus an invariant  $E(t)$
 cannot show {\em any effects} of  the two-way coupling.
It is important to note that it is appropriate to use forced isotropic turbulence to study the dispersion of bubbles or particles in {\em one-way coupling }(e.g. \cite{wang-maxey1993, katz2007}).
 }

\subsubsection{Deformable  bubbles with $d< \eta$}
\label{sec:deformicrob}
Performing DNS of turbulent dispersed multiphase flows to resolve the shape deformation of millions of bubbles or droplets with $d< \eta$  is beyond the capabilities of current parallel supercomputers.
However, it is feasible to use the point particle approach combined with a phenomenological
subgrid model to calculate the shape deformation of the dispersed phase.
 The first study (and  the only one at present) that followed that approach is by  \cite{Lohse2017} and thus will be described here in some detail.
 \cite{Lohse2017}  performed DNS  to study the flow of deformable sub-Kolmogorov bubbles dispersed in a turbulent Taylor-Couette flow.
They used two-way coupled point-particle
approach and were able  to simulate approximately $10^5$  continuously deforming bubbles.
The density ratio, $\hat{\rho}$, of the bubble  gas density to that of the liquid  was  $10^{-3}$, the viscosity ratio
$\hat{\mu}$ was  $10^{-2}$   and the volume fraction of the bubbles
was $\Phi_v=10^{-3}$.
\black{The  study focused on the effect of deformability of bubbles on  the reduction of the  torque required to rotate the inner cylinder at a prescribed angular velocity.
 The reduction of the  required torque is calculated by comparing the average shear stress at the rotating wall
 for the bubble-laden flow with that of a single-phase flow.}
 The rotation rate of the inner cylinder was quantified by the inner cylinder Reynolds number, $Re_i= r_i \omega_i(r_o-r_i)/\nu$, where $r_i, r_o ,  \omega_i  \  \rm{and}\  \nu $
 are respectively the inner and outer cylinders  radii, the angular velocity of the inner cylinder, and
 the kinematic viscosity of the carrier fluid. Two cases were simulated with  $Re_i= 2.5\times 10^3\  \rm{and} \ 8\times 10^3$.
 \nl
 The bubble shape was  assumed to be  at all times a tri-axial ellipsoid described by a symmetric, positive definite second-rank
 tensor \bS   which satisfies the condition  \bSinv $ : \xvec \xvec = 1$,
 where $\xvec$ is the position vector of any point on the ellipsoid surface relative to its center. The time rate of change of \bS is described by  the phenomenological equation of \cite{Maffettone1998} which was originally developed for liquid droplets:
 \be
 \frac{d  \BS}{dt} - \Big[ \BOM \cdot \BS    - \BS \cdot \BOM \Big]   =  -\frac{f_1}{\tau}\ \Big[  \BS- {\rm{g}}(\BS)\BI  \Big] +
   f_2 \Big[  \BEE \cdot \BS  +    \BS \cdot \BEE   \Big] \ ,
   \label{equ:sequ}
 \ee
 where $\tau= \mu R/\sigma$ is the interfacial time scale,
$\mu$ is the dynamic viscosity of the carrier fluid (liquid), $R$ is the radius of the equivalent undeformed  spherical bubble, and  $\sigma$
is the surface tension.  $\BEE$   and   $\BOM$ are the strain rate and rotational rate tensors respectively. $\BI$ is  the second rank unit tensor.
  \nl The LHS of (\ref{equ:sequ}) is the Jaumann corotational derivative  (\cite{gurtin2010}) which is frame-invariant and depends on  $\BOM$.
  Equation(\ref{equ:sequ}) states that
 the temporal evolution of the {\em shape tensor}  \bS is governed by
  two competing phenomena:   the interfacial tension (first term on the RHS) which attempts to restore
the initial spherical shape,  and the drag exerted by the motion of the ellipsoid  (second term on the RHS) while preserving the initial volume.
The positive dimensionless coefficients  $f_1$  and $f_2$ are functions of the viscosity ratio $\hat{\mu}$.
The function ${\rm{g}}(\BS)$  is introduced to preserve the bubble volume and is proportional to the ratio of the third invariant of $\BS$
to the second invariant of $\BS$. The  derivation of (\ref{equ:sequ})  is given by  \cite{Maffettone1998}.
\nl
Time integration of (\ref{equ:sequ})  leads to
 three eigenvalues of \bS which equal  the squares of the three semi-axes of the ellipsoid, and three eigenvectors that  provide the orientations of the semi-axes.
 Equation(\ref{equ:sequ}) has been validated experimentally by \cite{guido2000} for a neutrally buoyant liquid droplet immersed in a viscous fluid subjected to uniform shear.
\nl  \cite{Lohse2017}  assumed  a small Capillary number which measures the relative importance of the viscous forces to surface tension forces at the small scale motion.
 $Ca = \tau/\tau_K = ( \eta \, \mu_f  ) /( \sigma \tau_K)  <<1$, where $ \mu_f $ is the dynamic viscosity of the carrier fluid and
 $\tau_K$ is the Kolmogorov time scale.
 \begin{figure}
{
\includegraphics[width=3.in]{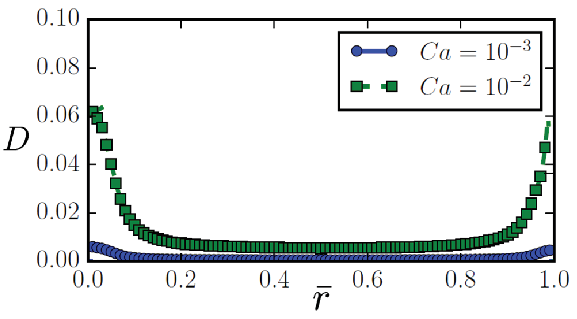}
}
\caption{Radial (wall-normal) profiles of azimuthally-, axially- and temporally-averaged deformation parameter $D$
of the sub-Kolmogorov length scale bubbles for two Capillary numbers, at $Re_i=2500$.
Source:   \cite{Lohse2017} with permission from the American Physical Society.}
\label{fig:TCbub2}
\end{figure}
\begin{figure}
{
\includegraphics[width=3.2in]{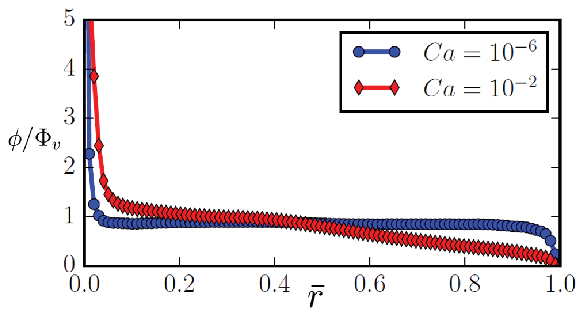}
}
\caption{Radial (wall-normal) profiles of azimuthally-, axially- and temporally-averaged local volume fraction
of the sub-Kolmogorov length scale bubbles, $\phi$,  normalized by the total volume fraction, $\Phi_v$, for two Capillary numbers, at $Re_i=2500$.
Source:   \cite{Lohse2017} with permission from the American Physical Society.}
\label{fig:TCbub}
\end{figure}
\noi
They further assumed that $Ca_{max}= 0.1$ to ensure that the deformed ellipsoidal bubble is nearly axisymmetric.
They used the drag and lift coefficients developed by \cite{njob2015} for {\em solid} ellipsoidal particles by
assuming that the bubble interface is contaminated with surfactants to satisfy the no-slip boundary condition.
\black{\cite{se-lohse2017} justified using the no-slip boundary condition for the bubble surface and the drag and lift forces for solid ellipsoids by setting  $Ca<<1$ and $\hat{\mu}=10^{-2}$;  hence the viscosity of the bubble gas was negligible, thus minimizing the drag due to the internal gas circulation.}
 The bubble acceleration equation accounted for the forces due to drag, lift, added mass and buoyancy.
\nl
\black{The DNS results show that an increase in the deformability of the sub-Kolmogorov  bubbles
enhances drag reduction due to a significant accumulation of the deformed bubbles near the rotating inner wall (Figs.\ref{fig:TCbub2} and \ref{fig:TCbub}).   In Figure \ref{fig:TCbub2} the dimensionless {\em deformation parameter }is $ D= (d_3-d_1)/(d_3+d_1)$ (\cite{Maffettone1998}), where
 $d_1 \  \rm{and}\ d_3$ are the lengths of the minor and major semi-axes of the ellipsoid, respectively.
A larger concentration of bubbles
near the driving wall indicates  that they are  effective in \black{suppressing  the plumes ejection process;  hence
  drag reduction is greater.  These plumes are unsteady vortical structures that detach from either the inner or outer cylinder in wall-normal direction (\cite{Lohse-plume-2015}).
 The plume ejection occurs  predominantly at the stagnation regions (along the walls) between the counter rotating rollers
where a negative pressure gradient normal to the wall is created.
  }
 The bubbles preferential accumulation is induced by  increased resistance to  the bubbles leaving the wall  in its normal direction. The increased resistance is due to the strong deformation of the bubbles near the rotating wall which makes them prolate (stretched along one axis) and oriented along the streamwise direction.}
\subsection{Bubbles of  size larger  than the Kolmogorov length scale, $d >\eta$}}
\subsubsection{Single bubble deformation and breakup in isotropic turbulence using LBM}
\cite{qian2006} studied the deformation and breakup of a single bubble in stationary isotropic turbulence ($20 \le Re_{\lambda} \le 35$)
using LBM with D3Q15 lattice (cf. section \ref {subsec:LBM}) and the BGK (\cite{bgk54})
form of the LB equation with single relaxation time.
A validation test was made for a non-turbulent flow by
 comparing the results of this  LBM  with those of FTM for  a 2D rising bubble and good agreement  was achieved (\cite{sankar2003}).
 \cite{qian2006}  defined Weber number as $We= \rho_l\langle \delta u^2\rangle d_e/\sigma$, where
$\delta u$ is the velocity difference over a distance equal to the bubble equivalent spherical diameter, $d_e$, and $ \langle..\rangle$ denote
averaging over space and time.  The results  show that prior to bubble breakup,  an increase
in the bubble surface area by at least $37\%$  occurs. 
They compared their results with the experimental data of \cite{risso1998} which were conducted in a microgravity environment.
Both the experiment and LBM  indicate that a Weber number can be identified below which breakup is not observed.
This Weber number is based on the statistics of the single-phase flow that would
exist in the absence of the bubble. In LBM,  this Weber number was
approximately 3.0.
\subsubsection{Deformable bubbles  with  $d >\eta$ in turbulent channel flow using FTM}
\label{sec:deformb}
\cite{tryggv1992}  introduced the {\em front-tracking method } (FTM) in which one set of the incompressible  Navier-Stokes and continuity equations is used for the whole computation domain, including the bubbles and the carrier liquid:
\bea
  \rho\frac{\pa \uvec}{\pa t} +  \rho \nabla \cdot \uvec\uvec &=& - \nabla P + \nabla \cdot \mu (\nabla\uvec +\nabla\uvec^T)
+ \sigma \int_F k_f \, \nvec_f \, \delta (\xvec-\xvec_f) dA_f \ , \label{eq:NSFT}\\
   \nabla \cdot  \uvec &=& 0 \ ,
\eea
where $\uvec$  is the velocity, $P$ is the pressure, and $\rho$ and $\mu$ and  are the
discontinuous density and viscosity fields, respectively. 	$\delta$ is a
three-dimensional delta function. $\sigma$ is the surface tension coefficient,
$k$ is twice the mean
curvature, and the subscript $f$ denotes the `front'. $\nvec$ is a unit vector normal to the front.
The integral is over the entire front such that it
 creates a force acting at the
interface but smooth along the front. $\xvec$  is the point at which
the equation is solved and $\xvec_f$  is the position of the front.
\nl
The moving interface (front) between the bubble gas and the surrounding liquid is represented by unstructured  mesh
that explicitly marks the position of the front.
The front mesh (or {\em marker}) points are advected by the carrier flow velocities, interpolated from the fixed Cartesian mesh.
As the front deforms, surface markers are dynamically  added or removed.
An indicator function $\Ivec(\xvec)$ which has the values  1 inside the bubble and 0
in  the carrier liquid is constructed from the known position $\xvec_f$. Since both $\rho$ and $\mu$ are constant within each fluid, their values at any point can be calculated using $\Ivec(\xvec)$:
$\rho(\xvec) = \rho_0 + (\rho_b -\rho_0)  \Ivec(\xvec) , $
and
$\mu(\xvec) = \mu_0 + (\mu_b -\mu_0)  \Ivec(\xvec) ,$
where the subscripts $b$ and $0$ denote the bubble and carrier fluids respectively.
\nl To avoid numerical instabilities associated with a sharp interface,
the front is
 given a  thickness of the order of several mesh cells.
In this narrow transition zone the fluid properties change
smoothly. The sharp delta functions in (\ref{eq:NSFT}) are approximated by smoother functions  with a compact stencil on the fixed Cartesian mesh.
At each
time step, after the front has been advected, the density and
the viscosity fields are reconstructed by integration of the
smooth grid delta	 function. The surface tension force (last term on the RHS of (\ref{eq:NSFT})) is then added to
the nodal values of the discretized Navier-Stokes equations.
More details are given by \cite{tryggv2006}.
\par \cite{tryggv2005}  performed DNS of a turbulent flow in a `minimum channel' at $Re_{\tau} = 135$.
 They used the front tracking method (FTM) described above
 to study the effect of 16 dispersed  bubbles on the wall shear stress. The nondeformed bubble diameter was $54$ wall units. The density ratio was restricted to
$\rho_b /\rho_0= 0.1$ and dynamic viscosity ratio  $\mu_b /\mu_0 =1$ which resulted in the kinematic viscosity
ratio $\nu_b/\nu_0 = 10$, a typical value for air bubbles in liquid water. Three Weber numbers, $We = 0.203, 0.270$  and $0.405$, were tested. The results show that deformable bubbles (with largest $We$ ) can lead to significant reduction
of the wall drag by suppression of streamwise vorticity. Less deformable bubbles, on the other hand,
are slowed down by the viscous sublayer and lead to a large increase in drag.
\black{The  reduction of  streamwise vorticity occurs by bubbles
moving parallel to the wall at a distance of about one
bubble radius between the bubble surface and the wall.
These bubbles move over the  streamwise vortices with a higher velocity than the advection velocity of the vortices. The passing of the bubbles forces the streamwise vortices closer to  the wall,  causing mutual cancelation between the streamwise vortices and the wall-bounded vorticity of the opposite sign. This results in a reduction of  the streamwise vorticity and a corresponding reduction of the
$\overline{uv}$  component of the Reynolds stress tensor.}
 \cite{tryggv2006} discuss the numerical difficulties that arise when using FTM with large density and viscosity ratios.
\par
\cite{tryggv2006b} performed DNS with FTM to study a bubbly vertical channel {\em downflow} at $Re_{\tau}= 127$.  The number of bubbles ranged from 18 to 72, and
diameter = $0.25 H$,  with an average volume fraction ranging from $ 0.015  \le \Phi_v \le 0.06$,
where $H$ is half the channel width.  The density and kinematic viscosity  ratios were respectively  $\rho_b /\rho_0 = 0.1$ and $\nu_b/\nu_0 = 10$.
The results show that the lift force on the bubbles near the wall caused them to concentrate in the core region and create  a bubble-free wall layer.
\par
\cite{tryggv2008} performed  DNS with FTM  to study the effect of deformability of 21 bubbles on their dispersion in a turbulent bubbly {\em upflow} in a vertical channel at friction Reynolds number, $Re_{\tau}=127$. The density and viscosity ratios were prescribed as those in \cite{tryggv2006b}. Two cases were studied for two values  of $E\ddot{o}tv\ddot{o}s$ number, $Eo = \rho_f\, {\rm g } d^2/\sigma$, which measures the ratio of gravitational forces to surface tension forces;
  $Eo = 0.45$  for nearly spherical bubbles and 4.5 for deformable bubbles.
\black{The lift force on a clean spherical bubble rising in a vertical shear flow is directed toward
the side where the fluid moves faster  past the bubble, in a frame of reference
moving with the bubble (\cite{tryggv2006c}) according to \cite{saffman65}.
In channels, where the fluid velocity is zero at the walls, {\em spherical } bubbles will therefore
move laterally toward the walls for upflow and away from the walls in downflow.}
 However, bubble deformation can reverse the sign of the lift force.  Furthermore, the  lift force on strongly deformable bubbles is weaker than that on  nearly spherical bubbles.
 Turbulent dispersion of deformable bubbles overcomes their tendency to concentrate in the core region of this upflow.
 \par
\black{
\cite{tryggv2013} performed DNS of {\em nearly spherical bubbles }in a vertical turbulent channel upflow similar to their earlier study (\cite{tryggv2008})
but at   higher Reynolds number,  $Re_{\tau}=250$, and 140 bubbles.  The bubble diameter was prescribed equal to 40 wall units = 0.08 of the channel width. }
\black{At statistically steady state, the weight  of the  bubble-laden liquid and the imposed pressure gradient  are balanced by the shear stress.
For the upflow in the channel,  as the bubbles migrate toward the wall (by the lift force) the average mixture density {\em in the core} increases
until the weight is balanced exactly by the pressure gradient. The shear and subsequently the lift force vanish in the core region, thus  ending
 the bubbles migration to the wall. As a result, the velocity gradient exists only in the wall region where the bubbles volume fraction reaches its peak (see Figs. 3 and 4 in  \cite{tryggv2013}).
The high concentration of the bubbles in the wall layer results in a significant reduction  of  the turbulence kinetic energy and its dissipation rate there.
  However, Fig. 5 of  \cite{tryggv2013} also shows that the {\em peak of the dissipation rate}
increases, as compared to single-phase flow, very close to the wall. The reason for that increase was not discussed in that reference.
}
\subsubsection{Deformable bubbles  with  $d >\eta$ in turbulent Taylor-Couette  flow using IBM-IPM}
\label{sec:bub-TC}
\black{
\cite{spandan2018} performed DNS  to study the effects of dispersed deformable bubbles, with  $d >\eta$, on drag reduction in a turbulent Taylor-Couette (TC)  flow using a hybrid IBM-IPM (section \ref{sec:intro}). { The surface of each dispersed bubble was discretized using an unstructured Lagrangian mesh.} The effect of the bubbles on the carrier fluid is accounted for via a volume-averaged  force computed on that mesh and then is transferred to the Eulerian mesh where the NS equations are solved.
The deformation of the immersed bubbles is computed via the IPM where the surface tension of a liquid-gas
interface is modeled using a triangulated network of elastic and torsional springs as described by \cite{spandan2017}.
Two cases were simulated with inner cylinder Reynolds number  $Re_i = 5\times 10^3\  \rm{and} \ 2\times 10^4$.
The number of the dispersed bubbles was 120, each with an initial  diameter $d_b\sim 14 \eta$ and $d_b\sim25 \eta$,
 for the low and high  $Re_i $ cases, respectively, and a global volume fraction $\Phi_v= 0.001$.
 Each bubble was initialized as a sphere with its surface
discretized with 1280 Lagrangian marker points for the low $Re_i$ case, and 2560 markers for the high $Re_i$ case.
Four values of  Weber number, based on the velocity of the inner cylinder, were considered: $We= 0.01, 0.5, 1, 2$.
These values  were assumed to be small enough to avoid bubble breakup or coalescence.
The  ratio, $\hat{\rho}$, of the bubble  gas density to that of the liquid  was  $5\times 10^{-2}$.
Bubble-bubble and bubble-wall collisions were modeled via elastic potential between the Lagrangian mesh nodes  and the center of the enclosing Eulerian cell.
\nl
The results show that for all four $We$ values the bubbles concentrate near the inner cylindrical wall. This is in contrast to the deformable sub-Kolmogorov bubbles which preferentially concentrate near the inner wall as  $We$  is increased (cf. section \ref{sec:deformicrob}).
\nl The drag reduction of the bubble-laden TC flow was derived by
\cite{sugiyama2008} as the sum of two terms : $DR_1 = \big(1- \frac{\langle \varepsilon_B  \rangle}{\langle \varepsilon  \rangle}\big)$
and $DR_2 = \big( \frac{\langle \fvec_b \cdot \uvec  \rangle}{\langle \varepsilon  \rangle}\big)$,
where ${\langle \varepsilon_B  \rangle}$ and  ${\langle \varepsilon  \rangle}$ are the mean dissipation rates of TKE per unit mass
of the bubble-laden carrier fluid and the single-phase flow, respectively. The volume-averaged source term in NS equations, $\fvec_b$, represents the two-way force per unit mass of the carrier flow due the dispersed bubbles, and $\uvec$ is the local fluid velocity. The results show that $DR_1$ dominates $DR_2$ and increases with increasing $We$.
The reason is that when the bubbles are more deformable, they are stretched along the streamwise direction similar to that of sub-Kolmogorov deformable bubbles (cf. section \ref{sec:deformicrob}).  The stretching reduces the projected surface area in the direction of the relative velocity which in turn lowers the  bubble Reynolds number,  leading to smaller wake, lower  TKE, thus smaller  ${\langle \varepsilon_B  \rangle}$ and lower $DR_1$.
}  
\section{DNS OF DROPLET-LADEN TURBULENT FLOWS}
\subsection{Droplets of  size smaller than the Kolmogorov length scale, $d<\eta$}
Examples of liquid droplets of diameter $d<\eta$ include rain droplets in the atmospheric boundary layer
and fuel droplets in the combustion chambers of modern aircraft jet engines.
 \cite{carl2001}  measured the size distribution of the liquid fuel droplets under realistic conditions in the combustion chamber
 of an aircraft jet engine
and their data show that the diameter of most droplets
  is smaller than  $\eta$ in the region downstream of the initial ligament breakup zone.
  \subsubsection{Nondeformable droplets with $d< \eta$}
\par
 \cite{mashayek98}
performed DNS with  the point-particle approach and  $96^3$ grid points to study droplet-turbulence
interactions in a homogeneous shear flow. He considered
both one-way and two-way couplings for non-evaporating
and evaporating droplets. The number of droplets was  $1.5\times 10^5$ and the diameter of non-evaporating
droplets varied between $0.2 \eta$  and $0.3\eta$.
The study showed that in the case
of non-evaporating droplets, the turbulence kinetic energy is
reduced and the flow anisotropy is increased due to the two-way
coupling. In the case of evaporating droplets he found
that the turbulence kinetic energy and the mean internal energy
of the carrier flow are increased due the  mass transfer of the droplet vapor to the carrier fluid.
 \par
  \cite{mclaugh1998} studied  the collision and coalescence  of mono-sized droplets in a turbulent channel using DNS with the point particle approach
 at a friction  Reynolds number, $Re_{\tau}= 116 $. The droplet diameter ranged from    $0.1 \eta$ to $0.5\eta$ and the initial volume fraction was in the range  $5.8\times 10^{-6 } \le \Phi_{v}\le 3.1\times 10^{-4}$. The ratio of the liquid density to that of the carrier fluid varied from  20 to  2000. The results showed that the droplet inertia was the dominant  factor in the collision mechanism.  The results also showed that the predicted collision rate agreed with the theory of \cite{saffman1956} for droplets with response time in wall units $\tau_d^+  <1$.
  \par
 \cite{bellan-jfm-99} performed DNS of a confined temporally-developing mixing layer with one layer laden with evaporating liquid droplets using the point particle approach and two-way coupling
 for mass, momentum and energy exchanges.
 The confining walls were treated as
 {\em frictionless  and adiabatic} for simplifying the boundary conditions for the droplets and ensuring the conservation of mass and energy.  The initial volume fraction of the droplets in the
 laden stream was $5.5\times10^{-4}$. The initial number of the mono-size droplets in the different cases varied from $4\times10^4$ to $7.3\times10^5$ and their initial diameter ranged from $115 \mu m$ to  $231\mu m$. The Reynolds number based on the vorticity thickness was  200, and the convective Mach number $M_c$ was 0.5. The initial temperatures of the gas and droplets  were respectively $350 K$ and  $325 K $. The results show that the turbulence kinetic energy  and the growth rate of the mixing layer were both
attenuated  monotonically  by increasing the mass loading ratios of the droplets.
 \par
\cite{bellan-jfm-05-1} extended the mathematical formulation  of  \cite{bellan-jfm-99} to account for multi-component chemical composition of the liquid droplets.
They examined the effects of the liquid
composition on the development of the vortical features of the flow, the vortical
state reached after the second pairing, and  the gas temperature and composition.
They concluded that the mixing layer growth and main rotational characteristics are unaffected by
liquid specificity; however, the global mixing is highly liquid-specific.
The  analysis of the vorticity budgets showed that
 the small-scale vortical activity increases with increased fuel
volatility.
 \par   \cite{kuerten2014}   studied the  evaporation and condensation of water droplets in a turbulent channel flow in zero gravity at $Re_{\tau} =150$ using DNS with the point particle approach and  two-way coupling of mass, momentum and heat between the two phases. The carrier fluid consisted of air and water vapor.
One of the channel walls was heated while the other was cooled. This created
a temperature gradient in the wall-normal direction  and also  a
non-uniform mean vapor mass fraction.
The objective of the  study was to analyze the effects of phase change on the global heat
transfer properties of the flow and on droplet motion and size distribution.
Both the densities of the air and water vapor were time- and space-dependent, but their sum remained invariant to satisfy the
zero divergence condition for the whole flow.
The details of the pseudo-spectral numerical method are given in \cite{kuerten2006}.
The results show that initially the droplets  migrate towards the
 channel walls due to turbophoresis (\cite{reeks1983}), thus increasing  the droplet concentration
in the vicinity of the walls. Simultaneously, evaporation and condensation result in  the
droplets growth near the cold wall and diminution  near the warm wall. This also
creates a gradient in water vapor concentration, directed from the cold to the
warm wall. After reaching a steady state, the droplet concentration and
mean droplet size become nearly constant.
Turbulent diffusion of water vapor generates a mean flux of water vapor from the warm
to the cold wall.  Consequently, conservation of water mass results in
 a net mass flux of the droplets from the cold to the warm wall.
\nl
The results show that at  steady state : (1) The heat transfer between the two walls for the droplet-laden flow, quantified by
the Nusselt number, is larger by a factor of 3.5 than that of the single-phase flow, and by a factor of 2.6 than that of a flow  laden
with solid spherical particles having the same diameter, response time, $\tau_p^+=\tau_p u^2_{\tau}/\nu$, and specific heat of the droplets. This augmentation of heat transfer (by droplets vs. solid particles)
is due to the latent heat of vaporization which reduces the droplet temperature near the hot wall and
the latent heat of condensation which increases the droplet temperature near the cold wall.
(2) The turbulence modulation of the carrier fluid by the droplets
is the same as the modulation by solid particles. The Reynolds shear stresses
and the TKE production are reduced in the wall region by the droplets. This finding is in contrast
to that of  \cite{mashayek98} who found that droplets evaporation enhanced the TKE production.
In the simulation of    \cite{kuerten2014} both evaporation and condensation occur due to the presence of the hot and cold walls
resulting in negligible net evaporation rate.  The mean droplet diameter was found to be smaller
near the warm wall than near the cold wall.  It is noted that nucleation of droplets and droplet
breakup were not accounted for in this study.
 \par
  \cite{kuerten2015} extended the DNS study of  \cite{kuerten2014} to include the effects of droplet collisions.
  The prescribed Weber number of the droplets was small such that coalescence between colliding droplets was negligible.
  The  droplets overall volume fraction was in the range $ 0.55\times10^{-4} \le \Phi_{vo} \le 2.2\times10^{-4}$, and the corresponding number of droplets varied from
  $ 0.5\times10^{6}$  to $ 2\times10^{6}$.
  The results show  that droplet collisions (i.e. four-way coupling)  cause a significant reduction (about 76\%) of the  maximum local concentrations
  of the droplets near the channel walls as compared to the  two-way coupling case with the same overall $\Phi_{vo}$.
  Regarding the dependence of droplet collisions  on   $\Phi_{vo}$,  \cite{kuerten2015}  stated that:
  ``Elghobashi's diagram (\cite{se-appl-sc-res94}) indicates that the demarcation line
between the two-way and four-way coupling regimes shifts toward lower volume fraction if the
Stokes number becomes higher. However, in the present work, the Stokes number is only 10 in wall
units, which shows that the effect of collisions on concentration in dilute flows is not limited to
very high Stokes numbers."
\nl
Two comments related to the above statement are made here to clarify the diagram of \cite{se-appl-sc-res94}:
 {\em (i)}
  The logarithmic-scale ordinate in the diagram is the Stokes number,  $St=\tau_p/\tau_K $.
   The diagram shows that the  demarcation line between the two-way and four-way regimes
   shifts toward lower volume fraction ($ < 10^{-3} $) for  $\tau_p/\tau_K \geq 0.7 $ since the maximum preferential accumulation  of solid particles in isotropic turbulence occurs at
     $\tau_p/\tau_K=1 $ (\cite{ferrante03}). In other words, particle collisions are  expected to start before the local concentration reaches its peak.
   {\em (ii)} If preferential accumulation  occurs in a particle-laden turbulent flow, then the abscissa of the diagram should represent   the {\em local volume fraction, $\phi_v$}
  (instead of {\em the overall  volume fraction, $\Phi_{vo}$}),
     to determine whether the regime {\em at a selected location } is two-way or four-way coupling.
\nl
The heat transfer results of   \cite{kuerten2015}  indicate that accounting for the droplet collisions (four-way coupling) reduces   Nusselt number by approximately $17\%$
 as compared to two-way coupling for the case with highest  $\Phi_{vo}$. This means that a reduction of $76\%$ in the maximum   $\phi_v$ near the wall resulted in {\em only}
  $17\%$ reduction in Nusselt number.
  \nl In order to explain this result  we should note here that the large increase in Nusselt number when inertial particles are present in the flow is caused {\em totally}
by  the direct convective heat transfer between the particles  and the carrier fluid due to their temperature difference.
 The reason for this {\em direct causality} is that the two-way momentum coupling between the particles and fluid {\em reduces} the turbulent shear stresses and TKE and  hence {\em reduces} the turbulent heat fluxes
 within the carrier fluid, e.g. quantities proportional to $<u_i T_f>$, where $u_i$ and $T_f$ are the fluctuations of
 the local fluid velocity and temperature (see Eqs. (20) and (21) in  \cite{kuerten2015} ).
 \nl
 Now, the $76\%$ reduction in the maximum $\phi_v$ near the hot wall reduces the total surface area of the droplets across which  heat is transferred from the hot fluid.
 Consequently, the fluid temperature (as well as the temperature difference between the fluid and droplets) near the hot wall is higher for the colliding droplets than in the case of no collisions.
Thus, the Nusselt number reduction for the four-way coupling case  is not as severe as that of $\phi_v$.
 \subsubsection{Deformable droplets with $d< \eta$}
 \label{sec:drop<eta}
 \cite{cristini2003} studied the deformation and breakup of  sub-Kolmogorov droplets in stationary isotropic turbulence.
 The objective was to  enhance the understanding of the droplet breakup process  beyond the phenomenological models of  \cite{kolmog-49} and \cite{hinze1955}.
At  the scale of these droplets, the viscous stresses,  $\BC T_{\mu}=(\mu/\tau_K)$,  dominate the inertial stresses, $\BC T_{\rho}=(\rho d^2/ \tau^2_K)$,
since $\BC T_{\mu}/\BC T_{\rho}= \eta^2/d^2$. Consequently, the local  velocity field in the vicinity of these droplets was assumed to be governed by the  Stokes flow equations.
Both the viscosity ratio of the droplet fluid to the carrier fluid 
and the corresponding density ratio were set equal to unity (neutrally buoyant droplets).
The droplets were treated as passive tracers with no effects on the carrier fluid.
Under these conditions, it was assumed that the trajectory  of a droplet center of mass is identical to that of the carrier fluid particle that coincided with it at an initial time.
The velocity field of the stationary isotropic turbulence,  at $Re_{\lambda}=54$, was obtained using a pseudo-spectral DNS method.
\nl
The velocity field around a droplet was obtained by iteratively solving, at each time step,  the boundary-integral equation for the Stokes flow
on a set of interfacial marker points that were distributed on the surface of the initially spherical droplet.
That velocity field is matched with the velocity of the turbulent flow near the droplet location via
linear expansion.
The boundary conditions for the local Stokes flow velocity field around the {\em deformable  droplets} were prescribed at the droplet interface  by the
continuous velocity and tangential stress and the discontinuous normal stress due to surface tension.
The droplet interface was adaptively restructured, between time steps,  to maintain uniform
resolution of the pointwise curvature with a prescribed accuracy
as described in detail by \cite{cristini2001}.
 The results of \cite{cristini2003}  included the history of  the deformation of two initially spherical droplets along their trajectories.
Depending on the local shear/strain rates, the droplet deformation stages included stretched ellipsoids and dumbbells that led to neck thinning and  pinch-off.
 \par
 \cite{meneveau2014} studied the deformation and orientation statistics of
sub-Kolmogorov ellipsoidal
droplets in isotropic turbulence.
Both the viscosity ratio of the droplet fluid to the carrier fluid 
and the corresponding density ratio were set equal to unity (neutrally buoyant droplets).
The droplets were treated as passive
tracers with no effects on the carrier fluid.
Each of the simulated droplets followed  the trajectory of the carrier fluid particle coinciding with its center at an initial time.
The stationary homogeneous isotropic  turbulent flow  was computed via  DNS at $Re_{\lambda}  = 185 \, \rm{and}\,  400$. The ellipsoidal droplet shape evolution was predicted via the phenomenological equation of \cite{Maffettone1998} described earlier  in section \ref{sec:deformicrob}. The prescribed initial droplet size was such that $d/\eta \leq 0.1$.
The trajectories of $7\times 10^3$ droplets  for the $Re_{\lambda}  = 185 $ case and
$15\times 10^3$ droplets  for the $Re_{\lambda}  = 400 $ case were computed.
The results show, as expected, that increasing the Capillary number, $Ca$, for a given $\tau_K$, the droplet deformation increases.
The  deformation of a  typical droplet may follow a sequence of oblate, prolate and then return to spherical shape.
A critical Capillary number was identified at which the droplet elongation along one or two directions becomes unbounded, which
should eventually lead to droplet breakup.
\par
 \cite{Lohse2016} studied the deformation and orientation statistics of
neutrally buoyant sub-Kolmogorov ellipsoidal
droplets in turbulent Taylor-Couette flow. They followed the same approach of  \cite{meneveau2014}, and  their own study of the sub-Kolmogorov bubbles (\cite{Lohse2017}) described above  in section \ref{sec:deformicrob}.  \nl
\cite{Lohse2016} named the approach of considering the droplet as a massless passive tracer
a {\em zero-way coupling}.  However, {\em zero-way coupling } means {\em no coupling}, and certainly this is not the case  of a tracer following the identical  instantaneous Lagrangian motion of a fluid particle.
A {\em tracer} or {\em passive scalar}  is a more appropriate name for this approach.
\nl
The droplet sizes were in the range of $   0.05 \le d/\eta \le 0.15$ even during the deformation. The DNS were performed for  two inner cylinder Reynolds numbers ($Re_i= 2500, 5000$), four different Capillary numbers ($Ca = \tau/\tau_K = ( \eta \, \mu_f  ) /( \sigma \tau_K)=0.05, 0.1, 0.2\ \rm{ and} \  0.3$), and two viscosity ratios ($\hat{\mu}= 1 \  \rm{ and} \  100$).
\nl  The statistical analysis of  the droplet deformation was performed using  the dimensionless {\em deformation parameter} $D$ defined above in section
\ref {sec:deformicrob}.
The results show that the maximum values of $D$ occur near both the rotating and stationary walls,  and as expected, $D$ increases with increasing the Capillary number, $Ca$.
However, the peak of the $D$ profile moves away from the wall with increasing  $Ca$. This is a result of the elastic collision model used for the interaction of the droplet with the wall.
The center of mass of a highly stretched droplet is displaced away from the wall as compared to that of a less deformed droplet.
\subsection{Droplets of size larger than the Kolmogorov length scale, $d >\eta$}
\subsubsection{Single droplet deformation and evaporation  in isotropic turbulence using LBM}
\label{subsec:LBM}
\cite{albernaz-2017}
used a {\em hybrid} lattice Boltzmann method (LBM) to study the deformation and evaporation of a single droplet in stationary isotropic turbulence. In their hybrid method,  the fluid density and velocity fields were obtained via LBM with D3Q19 lattice and  multirelaxation-time (MRT) collision operator (\cite{Ginzburg2002}), but the internal energy conservation  equation was solved by the finite-difference scheme of \cite{lbmhybrid2003}.
In the D3Q19 lattice,  ``D3" denotes three-dimensional flow and  the ``Q" refers to the first author of the paper \cite {qian-1992}.
The number 19  indicates that a  fluid point at the center of the cubic lattice interacts with the 18  neighboring points (=12 points of intersection of the three midplanes with the edges of the cube + 6 points of intersection of  the three midplanes at the six surfaces of the cube).  A fluid point at the center of the cube has 18 possible velocity directions plus a zero velocity.
The internal  energy equation contained a correction term proportional to the difference between the mean pressure of the domain and the
initial reference pressure. The correction is needed for conditions close to the critical point where  fluctuations of thermodynamic
properties occur.
\nl Forcing at low wavenumbers using the  method of  \cite{kareem2009} was applied at every time step to generate a statistically stationary velocity field for $ 73 \le Re_{\lambda}\le 133$.
\nl The pseudo-potential  method of  \cite{shan1993} and \cite{kuper2006} was used to simulate the droplet in the LBM.
The liquid  hexane droplet was surrounded by its vapor as the carrier fluid.
The  interface between the liquid and vapor was considered as
a thin transition layer of finite width (several nodes of lattice) where the density changes smoothly from one phase  to the other.
The ratio of the liquid density to that of the vapor was $ \approx 10$.  Also,  the ratio of the liquid dynamic viscosity to that of the vapor was $ \approx 10$,
since both liquid and vapor had identical kinematic viscosity.   The surface tension, $\sigma$, was calculated via the Young-Laplace equation which relates the pressure jump across the interface to the product of  $\sigma$ and the local curvature  (\cite{landau-lifshitz}).
\nl The initial temperature of both the  liquid and vapor was prescribed equal to $0.9 T_{critical} $ of hexane.
The initial droplet diameter, $d_o$, ranged from $50 \eta$ to $80 \eta$, which corresponds to  the range of $2.4\lambda$ to $3.8\lambda$.
\nl The effect of surface tension on the droplet deformation  was  studied by varying the Ohnesorge number,
$Oh= \mu_{\ell}/  \sqrt{  \rho_{\ell}   \sigma  \, d }= \sqrt{We_{\ell}}/Re_{\ell}$,  in the  range  $ 4.2\times 10^{-3}  \le   Oh \le  6\times 10^{-3} $,  where $\mu_{\ell}$ and   $ \rho_{\ell}$ are the dynamic viscosity and density of the droplet.
\begin{figure}
{
\includegraphics[width=5.in]{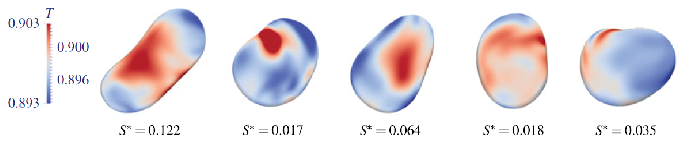}
}
\caption{Temperature distribution over the droplet surface for different values of the  deformation  parameter $S^*= (S-S_0)/S_0$,
where $S$ is the instantaneous area of the droplet surface, and $S_0$ is the equivalent surface area of a sphere whose volume is
identical to that of the deformed droplet.
Source:  \cite{albernaz-2017} with permission from Cambridge Univ. Press.}
\label{fig:dropdeformtemp}
\end{figure}
Some interesting results of this study  are:
 \begin{enumerate}[label=(\alph*)]
      \vspace{-3mm}
 \item For a fixed $Re_{\lambda}$, increasing $d_o$ increases the kinetic energy of the carrier fluid and reduces the kinetic energy of the droplet since the total kinetic energy is a function of the fixed $Re_{\lambda}$.

     \item The droplet deformation  increases  with increasing $d_o$ due to the increase of Weber number.

\item Reducing the surface tension increases the fluctuations of the thermodynamic properties, thus increasing the evaporation rate.

\item  At the droplet surface,  low temperature regions are associated with
stronger curvature whereas higher temperature occurs in flatter surface regions (Fig.\ref{fig:dropdeformtemp}).

\item Droplet volume fluctuations are correlated with vapor temperature fluctuations.
Strong correlations occur between positive temperature fluctuations and vapor
condensation.
\end{enumerate}

\subsubsection{Single droplet deformation in a turbulent channel flow using PFM}
\cite{scarb-soldati2013} used the Phase Field Model (PFM), section \ref{sec:intro}, of \cite{gurtin1996}  to study the deformation of a single droplet released  in a fully developed  turbulent channel flow at $Re_{\tau}=100$.
The ratios of  the liquid density and viscosity   of the droplet to those of the carrier fluid were prescribed equal to unity.
As described earlier in section \ref{sec:intro}, the forcing term representing the surface tension effects was added to the NS equations.
The modified NS equations were solved together with the phase field transport equation of \cite{cahn1959}.
The initial droplet diameter was $d_o=0.8H$, where $H$ is half the channel height, and    $  7.7 \le d_o/\eta \le 16.7$.
The Weber number, $ We =(\rho u^2_{\tau} H)/\sigma$,  was in the range $5.3\times 10^{-3}\le We\le 42.4\times10^{-3}$.
For that range of $We$, the mass loss at the end of the simulation  varied from $4\%$ to $14\%$.
The results show that the  TKE reaches its minimum value at the interface.
The  vorticity peak occurs at  a distance of  $\sim d/4$ from  the droplet interface, and the magnitude of that peak  decreases with decreasing $We$
(see  Fig. 5 in \cite{scarb-soldati2013}).
\par
\cite{scarbolo2013} compared the performance of  PFM  with that of LBM  (described in subsection \ref{subsec:LBM} above)
in simulating  the deformation of  a single 2D cylindrical  droplet in  simple  shear flows.
The comparison showed that the PFM's computational
cost  is almost three times higher than that of LBM. However,  the results of the  PFM appear
to be more accurate in that the spurious currents of the local kinetic energy created along the droplet
interface were smaller in PFM than in LBM by two orders of magnitude (see Fig.1 in \cite{scarbolo2013}).
\subsubsection{Coalescence and breakup of large droplets in turbulent    channel flow using PFM}
\cite{scarb-soldati2015} used the above described PFM in simulating  droplets with
initial number  $N_0=256$, and volume fraction $\Phi_v=0.054$, in a DNS of a  fully developed turbulent channel flow at $Re_{\tau}=150$. The objective was to study the interactions between the droplets.
The ratios of  the liquid density and viscosity   of the droplet to those of the carrier fluid were prescribed equal to unity.
The initial droplet diameter $d_o=0.4H$,  and $  15.9 \le d_o/\eta \le 37$,
where $H$ is half the channel height.
The interface thickness, $\xi$, was a constant prescribed via Cahn number, $ Ch= \xi/H= 0.0185$, which resulted in $0.36 \le  \eta/Ch \le 0.84$, thus minimizing the effects of the smallest eddies on distorting the interface.
The mass loss at the end of the simulation  varied from $2\%$ to $10\%$ for the range of $0.18\le We\le2.8$.
The results show that droplets, under the selected conditions and fluid properties,  migrate away from the wall toward the channel center.
The study identified two regimes of droplets interactions based  on the Weber number.
For $ We<1$,  the relatively large surface tension prevents droplet breakup,  and  allows   coalescence events to prevail.  Eventually, for  $t^+>2000$ (in wall units), the number of the merged droplets becomes  $ < N_0$, and their separation distances increase resulting in diminished collisions.
\nl In contrast, for $We>1$,  the droplets breakup and coalescence processes  occur simultaneously  during an early transition period.  This is followed at  large $t^+$ by  a dynamic equilibrium state at which the number of droplets reaches an asymptotic value which is about an order of magnitude larger than that for the case of $We<1$.
\par
\cite{scarb-soldati2016} performed DNS with the same flow conditions and fluid properties  of the above described study of \cite{scarb-soldati2015}
to investigate  turbulence modification by dispersed deformable droplets. The results show that for $We>1$ the  normalized wall shear stress or  friction coefficient, $C_f$, for the channel flow is not affected by the deformed droplets and its temporal development is nearly the same as that of the single-phase flow. However, for $We<1$, the temporal development of   $C_f$  shows a gradual increase at early times,  reaching a peak at  $1000\le t^+\le2000$, followed by a  gradual reduction. The physical explanation for these observations was not provided.
  \par
\begin{figure}
\vspace{-5mm}
{
\includegraphics[width=6.5in]{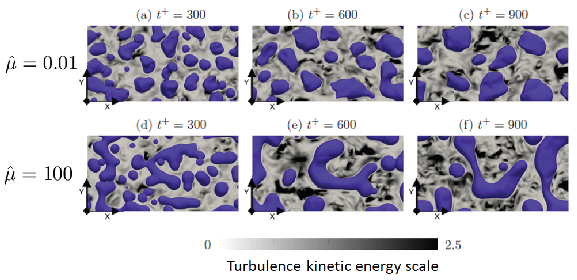}
}
\caption{ Temporal evolution of droplets for $We = 0.75$  and for two viscosity ratios
 (a-c)  $\hat{\mu} =0.01$ and (d-f)   $\hat{\mu} =100$.
Each panel refers to a given time instant
($t^+ = 300, 600,  \rm{and} \ 900$). Iso-contours of TKE computed on a
plane passing through the channel center are  shown in grey scale.
Source:   \cite{roccon2017} with permission from The American Physical society.}
\label{fig:dropbreak-A}
\end{figure}
\begin{figure}
{
\includegraphics[width=6.5in]{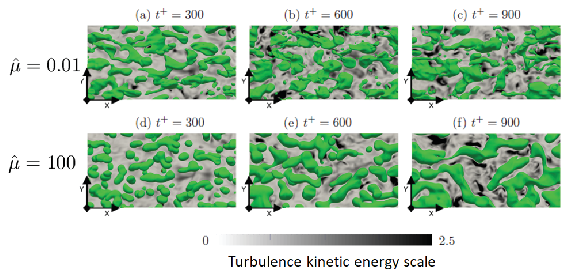}
}
\caption{ Temporal evolution of droplets for $We = 3$  and for two viscosity ratios
 (a-c)  $\hat{\mu} =0.01$ and (d-f)   $\hat{\mu} =100$.
Each panel refers to a given time instant
($t^+ = 300, 600,  \rm{and} \ 900$). Iso-contours of TKE computed on a
plane passing through the channel center are  shown in grey scale.
Source:   \cite{roccon2017} with permission from The American Physical society.}
\label{fig:dropbreak-B}
\end{figure}

  \cite{roccon2017} extended the DNS study of \cite{scarb-soldati2016},  described above,  by relaxing  the restriction   of unity viscosity ratio
   to examine the effects of varying the viscosity of the droplet.
Five different values of the dynamic viscosity ratio, $\hat{\mu} =0.01, 0.1, 1, 10,100$,
and three values of Weber number, $We = 0.75, 1.5, 3$, were studied providing a total of 15 test cases.
\nl
The initial number of droplets was $N_0=256$, at a volume fraction $\Phi_{vo} = 0.183$ and   initial droplet diameter  $= 0.6H$.
\nl The results show that for all test cases,  the deformable droplets migrate away from the wall  and reduce  the wall friction slightly as indicated by an increase of  the average mean velocity in the central zone of the channel by $\sim 2\%-4\%$.
\nl
\black{Qualitatively, the results  show, as expected,  that increasing  the droplet viscosity or surface tension  decreases the breakup
rate.  For the case of the highest surface tension, $We = 0.75$, droplets coalescence rate overtakes their  breakup rate  for all values  of $\hat{\mu}$, resulting  in
a gradual reduction of the number of droplets (Fig.\ref{fig:dropbreak-A}) which reaches after time $t^+=1000$  an asymptotic value of about $0.04 N_0$.}
\noi\black{For the lowest surface tension, $We = 3$, the effect of varying $\hat{\mu}$ becomes more pronounced (Fig.\ref{fig:dropbreak-B}). For $\hat{\mu} \le 1$, the breakup rate increases and    the asymptotic value of
droplets number is about $0.4 N_0$.  For $\hat{\mu} = 10$, the  breakup rate decreases and the droplets number reaches $0.1 N_0$.
For  $\hat{\mu} = 100$, the coalescence rate prevails and the droplets number diminishes to $0.01 N_0$.
\nl These results show  that lowering the droplet viscosity (relative to that of the carrier fluid), at a fixed surface tension,  enhances the droplet deformation and the eventual breakup.}
\noi
The results also show, as expected,   that the mean curvature of the interface between the droplet and the carrier fluid depends on   $\hat{\mu}$.
 The interface is defined as the isosurface of  the scalar phase field function $\phi(\xvec,t)=0$, and its mean curvature is
 $ \kappa= \nabla \cdot \big( - \frac{\nabla  \phi}{| \nabla \phi|} \big)$  (\cite{sun2007}).  The case of lowest surface tension, $We = 3$,
and smallest $\hat{\mu} =0.01$ resulted in strong curvature and breakup leading to the creation of small droplets (Fig.\ref{fig:dropbreak-B}). In contrast, for the same surface tension and $\hat{\mu} = 100$,  large elongated droplets with relatively small  curvatures were created (Fig.\ref{fig:dropbreak-B}).
\subsubsection{Dispersion of liquid droplets in isotropic turbulence using LBM-PFM}
\label{sec:LBM-PFM}
\begin{figure}
\vspace{-5mm}
{
\includegraphics[width=4.5in]{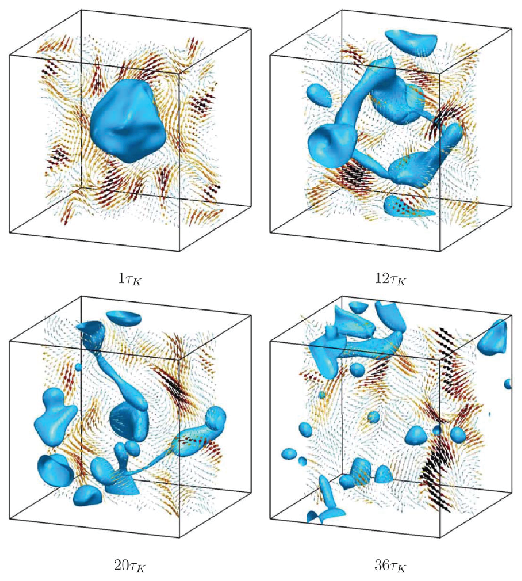}
}
\caption{Iso-surfaces of the scalar phase field $\phi=0$ representing the surface of the dispersed liquid droplet and velocity vectors at different time instants relative to the Kolmogorov time scale, $\tau_K$, for the case with $\eta=1$  [Lattice Unit], viscosity ratio $\hat{\mu}=1$. density ratio = 1,  and the Capillary number $Ca=0.1$.
The dispersed phase volume fraction is $\Phi_v= 0.03$. The initial  single droplet is placed in the isotropic turbulence at $t=0$.
Source: \cite{komrakova2015}  with permission from AIChE.}
\label{fig:kormakova-droplet}
\end{figure}
\cite{komrakova2015}  used the {\em free-energy} LBM of \cite{swift1996} in which the diffuse interface evolves naturally according to
thermodynamics. This {\em free-energy} LBM   is a hybrid of LBM and PFM.
\noi
 Two probability density functions,   $f^n_i(\xvec,t), n=2$,  described earlier in section \ref{sec:intro} are used:  one  to solve the continuity and  NS equations for the carrier fluid,
and the other to solve the Cahn-Hilliard equation (\cite{cahn1959}) described in section \ref{sec:intro}.  A single relaxation time collision operator (\cite{bgk54}) was used in the solution.
\nl
The D3Q19 lattice  (section  \ref{subsec:LBM}) was used to perform DNS of  stationary isotropic turbulence
generated by
 the linear forcing method of \cite{lundgren2003}.
The periodic  cubic computational   domain contained a maximum of $1000^3$ dimensionless  {\em lattice units} [lu]
where the Kolmogorov length scale $\eta \sim$1-10 [lu].
The Reynolds number $Re_{\lambda}$  could not  be prescribed  as an input parameter since the velocity $u_{rms}$ is
not known a priori. Only three parameters were used to prescribe the turbulent two-phase flow: the droplets volume fraction,$ \Phi_{vo}$, the viscosity ratio, $\hat{\mu}$,
and the Capillary number,   $Ca = \tau/\tau_K = ( \eta \, \mu_f  ) /( \sigma \tau_K) $, where $\mu_f$ is the dynamic viscosity of the carrier fluid.
The density ratio of the droplet to that of the carrier fluid was set to unity, and the viscosity ratio was $ 0.3 \le \hat{\mu}  \le 1   $.
\nl
The DNS starts at $t=0$ with a single droplet placed  in the computational domain with a volume fraction range  $0.001 \le  \Phi_{vo} \le  0.2$.
The initial droplet diameter was  in the range of 20$\eta$-30$\eta$.
The droplet breaks up due to the turbulent stresses in the carrier fluid
as shown in Fig.\ref{fig:kormakova-droplet}.  The Reynolds number was computed from  the DNS  results for the case  shown in
Fig.\ref{fig:kormakova-droplet} as
$Re_{\lambda}=42$  ( \cite{komrakova2017}).
\noi The study points out to the following  limitations of the DNS using LBM-PFM:
(a)
Coalescence occurs when  the interfaces of multiple droplets
 occupy the same computational cell. Thus, in order
to suppress unphysical coalescence, it is necessary to resolve
the liquid film between the droplets which requires prohibitive  mesh
refinement (\cite{shardt2013}).
(b) Dissolution of small droplets is  an inherent property of the numerical method (\cite{keestra2003}).
The dissolution rate increases as the droplet size decreases. To minimize the dissolution,  it is necessary to increase the resolution by keeping
    the droplet diameter in the range of 20$\eta$-30$\eta$.
(c) It is not possible to obtain an accurate  TKE
spectrum of a two-phase turbulent flow. It is known that LBM
is prone to generation of spurious currents due to   discretization  of the velocity space. The order of magnitude of
the spurious currents can be the same  as that of  the actual
velocity field. In addition, the spurious currents
appear within the diffuse interface and interact with the small scale motion
leading to a significant unphysical
energy gain at high wavenumbers.

\subsubsection{The interaction between  3130 fully resolved droplets and isotropic turbulence using VOF}
\noi
\cite{doddferr2016} performed DNS of decaying isotropic turbulence, with an initial $Re_{\lambda} = 83$,  laden with 3130 nonvaporizing droplets with $d \approx  20 \eta   \approx  \lambda$.
The ranges of the  density and dynamic viscosity ratios were   $1 \le \hat{\rho} \le 100$ and $1\le \hat{\mu} \le 100$, respectively. The Weber number based on the r.m.s. velocity of  the carrier fluid  was in the range   $0.1 \le  We_{rms} \le 5$. The volume fraction of the droplets was $\Phi_v=0.05$ and the mass fraction ranged from 0.5 to 5.
\par
Before discussing the results of the simulations,  it is worth describing the novel method developed by \cite{doddferr2016}  for solving the Poisson equation for the pressure in an incompressible immiscible  two-fluid flows {\em with large density and dynamic viscosity ratios}. The method is described in detail by \cite{doddferr2014}. It is well known that the numerical solution of NS equations of two-fluid flows with nonuniform density requires solving a {\em variable-coefficient}  Poisson equation for the pressure in the form
\be
\nabla \cdot \Big( \frac{1}{\rho^{n+1}}  \nabla p^{n+1}   \Big) =\frac{1}{\Delta t} \nabla \cdot \uvec^* \ ,
\label{eq:poisson}
\ee
where $ \uvec^*$ is the approximate fluid velocity at time step $n+1$.
Solution of  (\ref{eq:poisson}) is conventionally performed using iterative multigrid methods
(\cite{zaleski1999})
or multigrid- preconditioned Krylov methods (\cite{sussman2000}). All these methods are much slower than the fast Poisson solvers (e.g. FFT).
However, the latter require
 the coefficient  of  $ \nabla p^{n+1}$  to be a constant, whereas the coefficient
$ \frac{1}{\rho^{n+1}}$  on the LHS of (\ref{eq:poisson}) varies in space and time. In order to overcome this
problem, \cite{dong2012} split the product inside the brackets of  (\ref{eq:poisson}) in a way to render
the variable coefficient of $ \nabla p^{n+1}$ a constant. The first step is to approximate the the product  on the LHS of (\ref{eq:poisson}) as
\be
\Big( \frac{1}{\rho^{n+1}}  \nabla p^{n+1}   \Big) \approx\frac{1}{\rho_0} \nabla p^{n+1} +
\Big(  \frac{1}{\rho^{n+1}}  -  \frac{1}{\rho_0}  \Big)   \nabla {p^*}
\ ,
\label{eq:poisson1}
\ee
where $\rho_0= min (\rho_1,\rho_2)$ and $p^*=2p^n   - p^{n-1}$.
Then, substitution of the approximation (\ref{eq:poisson1}) into (\ref{eq:poisson})
results in:
\be
  \nabla^2 p^{n+1}   =   \nabla \cdot \Big[ \Big( 1-\frac{\rho_0 }{\rho^{n+1} }\Big)
   \nabla p^*\Big] +   \frac{\rho_0}{\Delta t}\nabla \cdot \uvec^* \  ,
\label{eq:poisson2}
\ee

\begin{figure}
{
\includegraphics[width=4.5in]{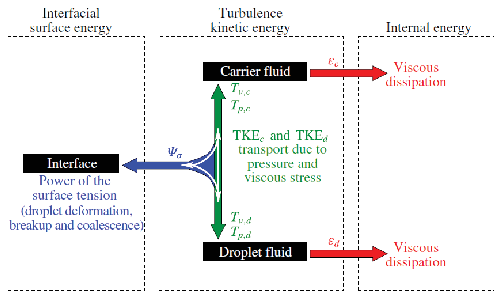}
}
\caption{Schematic showing TKE exchanges between the droplets and carrier fluid turbulence.
The three bounding rectangles from left to right  represent  the interactions between interfacial
surface energy, TKE, and internal energy. The blue arrow represents the two-way exchange between TKE
and interfacial surface energy  by the power of the surface tension, $\Psi_{\sigma}$.
The green arrows denote the transport of TKE between the two
fluids while  exchanging   TKE
for surface energy via $\Psi_{\sigma}$. The red arrows represent the transformation of  TKE of the carrier
fluid and droplet into internal energy by viscous dissipation.
[ From Fig. 5 of \cite{doddferr2016} with permission from Cambridge Univ. Press. ]}
\label{fig:TKEschematic}
\end{figure}
\begin{figure}
{
\includegraphics[width=6.in]{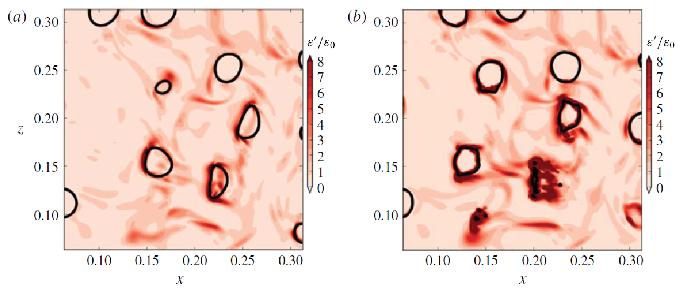}
}
\caption{Comparison of the viscous dissipation rate of cases E in (a) and F in (b).
Both cases have the same  dynamic viscosity ratio  $\hat{\mu}= 10$,
but case E has a density ratio $\hat{\rho}= 1$ , whereas case F has
$\hat{\rho}= 100.$
Source: Fig. 12 of \cite{doddferr2016} with permission from Cambridge Univ. Press. }
\label{fig:epsilonEF}
\end{figure}

\noi which can be solved using fast Poisson solvers.  \cite{doddferr2014} compared the accuracy and speed of solving (\ref{eq:poisson2}) for several canonical two-phase flows, at  density and dynamic viscosity ratios of values up to $10^4$,
  with that of  (\ref{eq:poisson}) and concluded that the new split method is at least an order of magnitude faster than that of  (\ref{eq:poisson}) for the same accuracy level.
  \nl Now, we continue the discussion of the main results of \cite{doddferr2016}, hereinafter referred to as DF.
  In order to explain the exchanges of TKE between the droplets and the carrier fluid, DF derived three TKE transport equations
  for the droplet phase, the carrier fluid and  the combined two-fluid flow [see Eqs. (B19), (B20), (B21) and (C8) in  \cite{doddferr2016}]. For the first two equations, the TKE decay rate  is governed by the sum
  of the viscous dissipation rate, $- \varepsilon$, the viscous power, $T_{\nu}$,  and pressure power, $T_p$. For the combined two-fluid, the TKE decay rate equals the sum of the viscous  dissipation rate and the power due to surface tension, $\Psi_{\sigma}$,
  which is the rate of work done by the surface tension forces on the fluid. $\Psi_{\sigma}$ can be a source or sink of TKE, depending on whether
  the total surface area of droplets  decreases  (e.g., coalescence)  or increases (e.g., breakup),  respectively.
  Figure \ref{fig:TKEschematic} shows a schematic
  of the TKE exchanges according to the three TKE transport equations.  The subscripts $c,d$ in the figure denote the carrier fluid and droplets respectively. \nl
    DF performed DNS for seven cases by varying $We_{rms}$, the density ratio, $\hat{\rho}$ or the dynamic viscosity ratio, $\hat{\mu}$.
   Increasing  $We_{rms}$ from 0.1 in case B  to  5 in case D
   showed that the number of droplets at the end of the simulation decreased relative to their initial number by about $20\%$  due to coalescence in B, and increased by about    $27\%$ in D due to breakup. In case B,  the power $\Psi_{\sigma}$ due to coalescence represents a source of TKE
   equal to about $50\%$ of the magnitude of the viscous dissipation rate.
\nl
Figure \ref{fig:epsilonEF} compares the viscous dissipation rate for cases E and F where both have the same $\hat{\mu}=10 $,  but their $\hat{\rho} = 1 \ \rm{and} \ 100$ respectively. The response time,  $\tau_d$, of the denser F droplets is 100 times that of the E droplets. Thus the F droplets generate higher  fluid strain rates near their surfaces than do the E droplets, and hence the dissipation rates are higher in F as shown in Figure \ref{fig:epsilonEF}.

\section{CONCLUDING REMARKS}
\black{
\begin{enumerate}
\item Significant progress has been made during the past 20 years in DNS of turbulent flows laden with  droplets or bubbles.
This progress is due to the continuing development of the numerical methods discussed above and the advances in supercomputer hardware and software. However, due to the complexity of these flows the experimental data needed to validate the DNS results are virtually nonexistent.
The needed experimental data should provide local instantaneous measurements of the velocity fields of both the carrier fluid and the dispersed phase
in addition to instantaneous images of shape changes of the latter.
\item Since Large Eddy Simulation will be used for the foreseeable future to predict turbulent multiphase flows at practical Reynolds numbers, accurate subgrid scale (SGS)  models need to be developed and validated by DNS results such as those described above. It is noted here that such accurate
    SGS models do not exist at present.
\item The   phenomenological equation of \cite{Maffettone1998} provides a powerful tool  for accounting of shape changes of deformable bubbles or droplets
which are smaller than the Kolmogorov length scale, $\eta$.
\item All the reviewed DNS studies of fully resolved droplets or bubbles, except that of   \cite{doddferr2016}, restricted the density and viscosity ratios of the two interacting phases in the range of  1 to 10.  Both the density and viscosity ratios  were equal to $10^2$ in the study of    \cite{doddferr2016}.
    Furthermore, \cite{doddferr2014} validated their numerical method with the analytical solution of the capillary wave of \cite{prosper1981} for density and viscosity ratios up to $10^4$,
    and with the experimental data  of \cite{beard1976} for a falling droplet  for a density ratio of $10^3$.
\item The PFM  provides qualitatively interesting results, however  it suffers from the following drawbacks :
(a) The large width of the interface region  (4-8 cells) leads to errors in the curvature (\cite{jacqmin1999}).
(b) The mass conservation is not satisfied  (\cite{yue2007}) as confirmed by
 \cite{scarb-soldati2013}  who showed that the mass loss at the end of the simulation  varied from $4\%$ to $14\%$,
 and from $2\%$ to $10\%$
  in the study of \cite{scarb-soldati2015}.
This inability to conserve mass renders the PFM quite inaccurate for cases involving vaporization or condensation as well as droplet motion
in highly vortical flows.
\end{enumerate}
}

\section*{DISCLOSURE STATEMENT}
 The author is not aware of any affiliations, memberships, funding, or financial holdings that
might be perceived as affecting the objectivity of this review.

\section*{ACKNOWLEDGMENTS}
The author thanks Professors William Sirignano, Andrea Prosperetti, Detlef Lohse, Alfredo Soldati and Antonino Ferrante for their helpful comments
on the first draft of this article.


\end{document}